\documentclass[aps,pra,twocolumn,english,preprint,10pt,nofootinbib]{revtex4}
\usepackage[T1]{fontenc}
\usepackage[latin9]{inputenc}
\usepackage{babel}
\usepackage{orcidlink}
\usepackage{graphicx,amsmath,amssymb,color}
\usepackage[normalem]{ulem}
\usepackage{amsfonts}
\usepackage{url} 
\usepackage{latexsym}
\usepackage{amsfonts}
\usepackage{amsthm}
\usepackage{mathrsfs}
\usepackage{natbib}
\usepackage{color,verbatim}
\usepackage{hyperref}

\newcommand{\be}{\begin{equation}}
\newcommand{\ee}{\end{equation}}
\newcommand{\bea}{\begin{eqnarray}}
\newcommand{\eea}{\end{eqnarray}}
\def\beq{\begin{equation}}
\def\eeq{\end{equation}}

\newcommand{\tev}{\,\, \mathrm{TeV}}
\newcommand{\gev}{\,\, \mathrm{GeV}}

\begin{document}

\title{Symbolic Classification-Enabled LHC Limits for BSM Global Fits}

\author{Shehu AbdusSalam\orcidlink{0000-0001-8848-3462}}
\email{abdussalam@sbu.ac.ir}
\address{Department of Physics, Shahid Beheshti University, Tehran, Iran}

\begin{abstract}
  Global fits of Beyond the Standard Model (BSM) physics often involve a two-way interplay between theory and experiment. Theoretical models provide guidance for experimental searches, while experimental results, in turn, constrain theoretical frameworks. A crucial aspect of this feedback loop is the direct inclusion of measurements and exclusion limits ``online'' global fits, i.e. during the parameter scans aspects of the global fits. However, incorporating the Large Hadron Collider (LHC) limits into such analyses has been computationally prohibitive, often due to time taken per parameter point exceeding the scales acceptable for global fit frameworks. In this study, we show that LHC limits can be incorporated ``online'' global fits by leveraging approximations derived from symbolic regression techniques. We utilize a dataset of ATLAS constraints from searches for electroweakino productions to derive a mathematical expression capable of classifying the phenomenological Minimal Supersymmetric Standard Model (pMSSM) parameter space as allowed or excluded. This is subsequently incorporated for making a global fit of the pMSSM to data, including the LHC Run-2 limits.

\vspace{2ex}
{\bf Keywords:} Symbolic Regressions, Large Hardron Collider, Beyond the Standard Model, Bayesian Inference
\end{abstract}

\maketitle

\section{Introduction}
\label{sec:Introduction}
The Large Hadron Collider (LHC) experiments provide an unprecedented volume of particle physics data, which are converted into measurements or exclusion limits on a wide range of physical observables. These results form the empirical foundation for evaluating theoretical physics scenarios beyond the Standard Model (BSM) of particle physics. Among the various strategies for confronting BSMs with data, global fits~\cite{0809.0284, 0904.2548, Cabrera:2009dm, Balazs:2009su, Lafaye:2004cn, Bechtle:2004pc, AbdusSalam:2022idz, AbdusSalam:2017uzr, Butter:2017aju, Costa:2019dbp, deBlas:2025xhe, GAMBIT:2025qto,Geoffray:2022vlh,2507.06191,2605.21594,Bruggisser:2021duo,1803.03252,2007.01296,1907.10619,1709.09414,1807.10660,2409.14934,1607.08517,1903.09578} have emerged as an important approach in the ongoing quest to uncover or exclude new physics proposals such as supersymmetry~\cite{Martin:1997ns,Quevedo:2010ui,Allanach:2024suz,AbdusSalam:2024vuo,ATLAS:2024lda, Constantin:2025mex}. This approach allows BSMs to be tested against a broad spectrum of observations.

The incorporation of LHC limits from direct searches for new physics into global fits poses a significant computational challenge. Most recasting and reinterpretation~\cite{2003.07868, 2203.10057, RIF:2025zgq} tools for the LHC limits require simulation pipelines, including the generation of Monte Carlo events, parton showering and hadronisation, detector simulation, event reconstruction, cross-section calculations, and the final application of analysis-level selections~\cite{Dercks:2016npn,Conte:2018vmg,Goodsell:2024aig,GAMBIT:2017qxg}. Executing this chain for every point in BSM parameter space is prohibitively time-consuming, making it impractical to embed directly within a global fit process. Consequently, most global-fit studies adopt an approximate strategy: the expensive collider constraints are applied \emph{a posteriori}, as a post-processing step on the parameter samples obtained from fits without the LHC limits~\cite{Allanach:2011ut, AbdusSalam:2012ir, Laa:2015aya, Barducci:2015zna}.

There are non simulations-focused tools which are based on simplified models scenarios~\cite{Kraml:2013mwa,Papucci:2014rja,1801.04847,1912.06374}. However, these are generally not well-suited for direct integration into comprehensive BSM global-fit frameworks because they are primarily tailored to simplified model spectra, which may not accurately capture the complexity of realistic BSM scenarios that often involve extensive cascade decays and multiple mass scales. This will be the case for a full phenomenological minimal supersymmetric standard model (pMSSM) scan. Tools like \texttt{SModelS}~\cite{Kraml:2013mwa,Ambrogi:2017neo,Ambrogi:2018ujg,Alguero:2021dig,MahdiAltakach:2023bdn,Altakach:2024jwk} are nevertheless suitable when addressing scenarios such as the electroweakino-pMSSM~\cite{Altakach:2023tsd,Constantin:2025bqp}, the explicit BSM example considered in this paper. 

Complimentary to the tools mentioned above, several studies have pursued using machine learning (ML) techniques to approximate collider constraints. A prominent example is the \texttt{SUSY-AI}~\cite{Caron:2016hib} software package which employs a Random Forest classifier~\cite{Breiman:2001hzm} trained on an extensive set of LHC Run-1 simplified-model and recast results in order to determine whether a given point in the pMSSM parameter space is allowed or excluded. Once trained, the model evaluates parameter points extremely rapidly. However, its training is restricted to LHC Run-1 analyses. Further, its software structure and dependencies makes its integration into global-fit pipelines cumbersome, limiting its usability. Its capability is also limited compared to \texttt{SModelS} from which LHC-limits based likelihoods can be extracted.

Other varieties of ML-based innovations have been explored across the broader global-fit pipeline, particularly in the domains of sampling, optimisation, and likelihood emulation; see e.g.~\cite{Feickert:2021ajf, Diaz:2024yfu,Diaz:2025sng,Chatterjee:2025gej,Hammad:2024tzz,Choudhury:2024crp,Baruah:2024gwy}. These include active-learning strategies for efficient exploration of high-dimensional parameter spaces, normalizing-flow-based density estimators and fast likelihood approximators, and more general simulation-based inference frameworks. These methods illustrate the growing role of machine learning for global fits studies.

In this study, we demonstrate that LHC direct search limits on BSM scenarios can be effectively approximated as mathematical functions of the model parameters. This approximation is achieved through the use of symbolic regression, a technique that aims to discover underlying symbolic expressions governing a dataset. This approach provides computationally efficient approximation for complex experimental constraints.

We explicitly apply the symbolic regression technique to an ATLAS dataset~\cite{ATLAS:2024qmx} on constraining pMSSM parameters with LHC Run-2 data. The resulting symbolic expression, which effectively encapsulate the experimental classification of parameter space, is integrated into a global fit framework for the pMSSM. This global fit was constructed by including only three example measurements, rather than a comprehensive set of all relevant experimental constraints. The rationale behind this specific construction is to demonstrate the main goal of this work: showing that symbolic classifications can readily approximate LHC limits, thereby enabling their incorporation ``online'' global fits in contrast to the \emph{a posteriori} post-processing approach. This implementation represents, to the best of our knowledge, the first instance of incorporating LHC direct search limits for new physics into a global fit of the pMSSM.

This article is arranged as follows. Section~\ref{sec:lhcSymbExpress} briefly describes the symbolic regression approach employed to approximate the ATLAS limits into mathematical formulae. The derived mathematical expression is then utilized to perform a global fit of the pMSSM. This fit incorporates both the LHC limits and three representative measurements: the supersymmetry contribution to the anomalous magnetic moment of the muon, the neutralino dark matter relic density, and the Higgs boson mass. For comparative purposes, and to assess the impact of the considered ATLAS limits, a second global fit is performed excluding the LHC limits. The results of these global fits are presented in Section~\ref{sec:globalFit}. We address the result of our studies in Section~\ref{sec.results} and conclude in the last section of the article.

\section{Symbolic expressions of LHC limits}
\label{sec:lhcSymbExpress}
This section outlines the methodology used to derive analytic approximations of experimental constraints. In particular, it details the application of symbolic regression to a representative set of ATLAS exclusion limits and presents an assessment of the accuracy and robustness of the resulting analytic expression.

The task of generating analytic expressions that reproduce the outputs of complex numerical computations falls within the domain of symbolic regression~\cite{Koza92}. In high-energy physics, this technique has been discussed in general terms and applied to specific phenomenological studies such as in~\cite{Udrescu:2019mnk,Butter:2021rvz,Bartlett:2022kyi,Koksbang:2023sab,Sousa:2023unz,Dong:2022trn,AbdusSalam:2024obf, Bahl:2025jtk, Bendavid:2025urn,AbdusSalam:2025the}. Symbolic regression proceeds by exploring the space of mathematical formulae to identify expressions that best describe a given dataset. Our goal is to obtain an accurate analytic representation of the ATLAS exclusion limits within the parameter-space region of the pMSSM. 

Within global fit analyses, such expressions serve as efficient approximation to the final output of the full, particle physics-based simulations pipeline. This yields substantial improvements in computational speed. For the classification task at the core of the study presented here, considerations of interpretability or algebraic simplicity are secondary; the main aim is to derive a mathematical formular that reproduces the ATLAS exclusion limits considered with high accuracy. 

A variety of algorithmic strategies for symbolic regression exist, including genetic programming and transformer-based approaches. Comprehensive reviews and comparisons of these techniques can be found in~\cite{defranca,DBLP:journals/corr/abs-1909-05862,DBLP:journals/corr/abs-2006-11287,DBLP:journals/corr/abs-2107-14351,cranmer2023interpretable,Burlacu:2020:GECCOcomp}. For the analysis presented in this letter, we employ the \texttt{Feyn}~\cite{Feyn} symbolic regression package, while noting that alternative software frameworks such as \texttt{PyOperon}~\cite{Burlacu:2020:GECCOcomp} and \texttt{PySR}~\cite{cranmer2023interpretable} could equally be used for such tasks if the limitations related to the highly multi-dimensional nature of the task are worked on.

The \texttt{Feyn} package is a supervised machine-learning tool for symbolic regression, inspired by Richard Feynman's path-integral formulation with line of reasoning similar to~\cite{Udrescu:2019mnk, Liu:2020omw, Liu_2022, aidescartes}. It searches over a large space of candidate mathematical expressions, each represented as a tree in which internal nodes correspond to mathematical operators (e.g.\ addition, multiplication, $\exp$, $\log$) and leaves correspond to input variables or constants. The fitness of each candidate expression is determined by its ability to reproduce the supplied dataset, typically quantified via a loss function such as the mean squared error.

In the following part of this section, we introduces the set of ATLAS electroweakino exclusion limits that forms the dataset which will be used for the symbolic regression task. 

\subsection{Summary of ATLAS limits considered}
This subsection provides a brief overview of the eight ATLAS Run-2 analyses incorporated into this study, which together probe complementary regions of the lightest supersymmetric particle (LSP) versus next-to-lightest supersymmetric particle (NLSP) mass plane. Certain subsets of the ATLAS electroweakino supersymmetry program are simplified models based~\footnote{Model independent results interpreted within simplified model frameworks.} and follows a strategy whereby multiple final-state topologies can constrain the same underlying production mode, depending on the variety of the intermediate boson decay channels appearing in the decay chain. The targeted scenarios include $\tilde{\chi}_1^\pm\tilde{\chi}_2^0$ and $\tilde{\chi}_1^+\tilde{\chi}_1^-$ pair production. The charginos decay through an intermediate $W$ boson (either leptonically or hadronically), while the neutral NLSP may decay via an on-shell $Z$ or Higgs boson. In the latter case, the Higgs boson is assumed to follow its SM branching ratios.

The ATLAS search channels included in this study, originally addressed in~\cite{ATLAS:2024qmx}, are summarised as follows. For each, wino/higssino LSPs and simplified model electroweakino-pMSSM scenario were assumed. The first targets ${\tilde{\chi}}_1^+{\tilde{\chi}}_1^-$ production with $W$-mediated decays, requiring exactly two opposite-sign electrons or muons, low hadronic activity, and substantial missing transverse momentum ($E_\mathrm{T}^\mathrm{miss}$)~\cite{1908.08215}. The second considers higgsino $\tilde{\chi}_1^\pm \tilde{\chi}_2^0$ production and $WZ$-mediated decays to $\tilde{\chi}_1^0$ whereby both the $W$ and $Z$ bosons decay leptonically~\cite{2106.01676}. The third is sensitive to chargino/neutralino production in final states with two boosted hadronically decaying bosons~\cite{2108.07586}. A fourth analysis targets $\tilde{\chi}_1^\pm \tilde{\chi}_2^0$ production where the $Z$ boson from $\tilde{\chi}_2^0$ decay and the $W$ boson from $\tilde{\chi}_1^\pm$ decay both decay leptonically~\cite{2204.13072}. The fifth probes the same production mode but with the neutralino decaying via an on-shell Higgs boson to $b$-jets, while the chargino undergoes a leptonic decay~\cite{1909.09226}. The sixth analysis considers electroweakino production with extended decay chains that give rise to final states containing four leptons~\cite{2103.11684}. The seventh search is optimised for ``compressed'' mass spectra, characterised by low-$p_\mathrm{T}$ leptons and large $E_\mathrm{T}^\mathrm{miss}$ arising from recoil against initial-state radiation~\cite{1911.12606}. The eighth is the disappearing-track search, targeting long-lived charginos that produce a short, terminating track in the inner detector accompanied by $E_\mathrm{T}^\mathrm{miss}$~\cite{2201.02472}.

In~\cite{ATLAS:2024qmx}, a set of 20{,}000 pMSSM electroweakino models was generated, of which 12{,}280 passed the ATLAS preselection criteria for event generation and reconstruction. These criteria include successful spectrum generation with a neutralino as the LSP, consistency with the LEP lower bound on the chargino mass, and a Higgs boson mass within the range 120~GeV to 130~GeV. For this, various software and analyses framework were used~\cite{GEANT4:2002zbu,ATLAS:2022yru, ATL-PHYS-PUB-2019-029,Heinrich:2021gyp}. For each model point, the combined likelihood evaluation yields a $\text{CL}_{\mathrm{s}}$~\cite{Read:2002hq} value, enabling a binary classification into excluded (CL$_\mathrm{s} < 0.05$) or allowed (CL$_\mathrm{s} > 0.05$) regions. In the ATLAS dataset, 2{,}263 of the 12{,}280 parameter points are excluded; excluded points are labelled as \texttt{exclusion = 0}, while allowed points are labelled as \texttt{exclusion = 1}.

The SLHA files corresponding to the pMSSM model samples -- together with the observed and expected CL$_\mathrm{s}$ values -- have been made publicly available~\footnote{\url{atlas.web.cern.ch/Atlas/GROUPS/PHYSICS/PAPERS/SUSY-2020-15/inputs/ATLAS_EW_pMSSM_Run2.html}}. This allows us to construct the dataset used for the symbolic classification task addressed in this study,
$$D = \{(x_i, y_i)\}, \qquad i = 1, \ldots, 12280,$$
where each $x_i$ denotes the vector of pMSSM parameters extracted from the SLHA file of model~$i$, and $y_i$ is a binary label indicating whether the model is excluded or allowed by the electroweakino searches. Specifically, $y_i = 0$ if the ``Final'' $\text{CL}_\mathrm{s} < 0.05$ , and $y_i = 1$ otherwise~\footnote{Since reducing the problem to a purely binary classification (``excluded/allowed'') could introduce artifacts, one may instead permit a small range of $\text{CL}_\mathrm{s}$ around 0.05 to avoid creating artificial features in the results.}. 

\begin{figure*}
    \centering
    \includegraphics[scale=0.35]{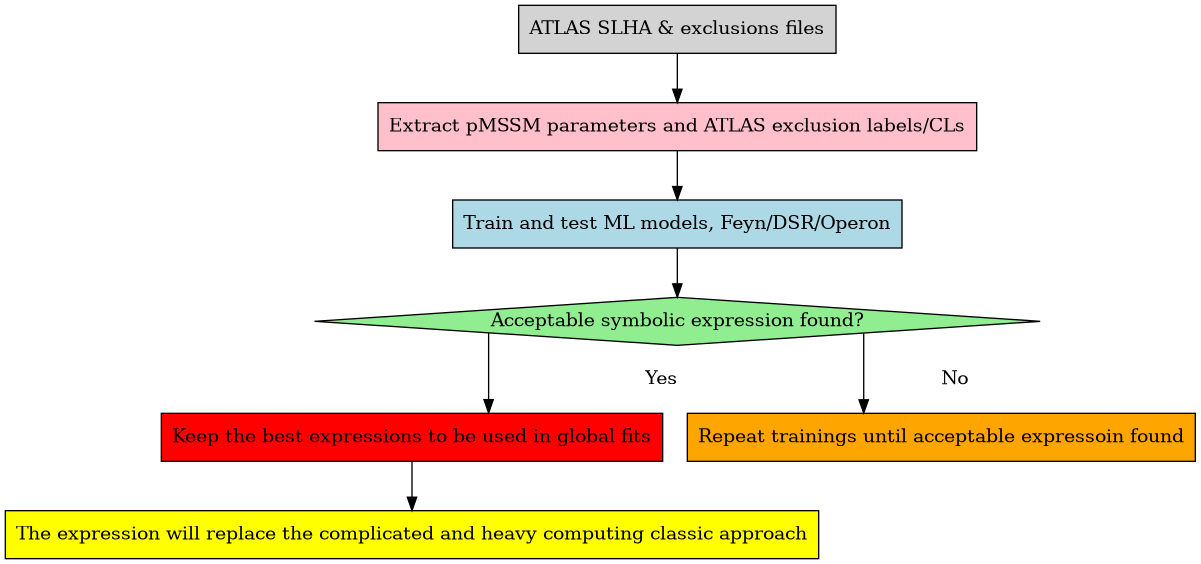}
    \caption{Schematic representation of the data extraction procedure and the objective of the symbolic classification task. \label{fig:SymbolicClassificationTask}}
\end{figure*}

Figure~\ref{fig:SymbolicClassificationTask} shows a schematic illustration for the computational pipeline of the symbolic classification task for deriving an analytic function $y(x)$ that approximates the ATLAS exclusion classifier $y_i$ across the pMSSM parameter space. Such an expression can subsequently be applied for the inclusion of the ATLAS limits ``online'' the global fit of the pMSSM. ``Online'' refers to applying the LHC constraint dynamically during the global-fit sampling, rather than as a post-processing step. 

\subsection{ATLAS limits as symbolic expressions}
In this subsection, we outline the symbolic-regression workflow implemented using the \texttt{Feyn} package to obtain analytical expressions directly from the dataset. The goal of the analysis is to derive a compact functional form capable of predicting the exclusion label $y(x)$ (the target or output to predict from data) from a set of the pMSSM parameters $x$ (the features or input variables for making predictions)~\footnote{A minimal working example of the code and the mathematical expression used for our analysis can be found via \url{https://gitlab.cern.ch/shabduss/symbolic-classification-enabled-lhc-limits-online-bsm-global-fits} and the dataset~\cite{abdussalam_2026_21262199}.}.

After loading the dataset, the relevant features were selected and the sample was randomly divided into training and test subsets, reserving 30\% of the data for testing. Because the exclusion labels are highly imbalanced, event-wise weights were computed using scikit-learn's \texttt{compute\_sample\_weight} function with the \texttt{class\_weight='balanced'} option. This assigns weights inversely proportional to the class frequencies~\cite{he2009learning, he2021weighting}, ensuring that both excluded and allowed points contribute comparably during the symbolic-classification optimization.

The symbolic-classification search was conducted via the \texttt{auto\_run} routine. This routine was configured for binary classification, employing a binary cross-entropy loss function. Model selection was guided by the Akaike Information Criterion (AIC)~\cite{AICprimer}, which inherently penalizes excessive complexity in the analytical expressions. Parallel evaluations were facilitated across 28 threads.

To ensure the convergence of the expressions towards a global optimum, a multi-stage iterative strategy was adopted. In this approach, the best model identified in a completed run was used as the initial population for subsequent runs. This ``warm-start'' technique enables the algorithm to progressively refine solutions by building upon previously discovered high-quality partial models, leading to more robust and accurate expressions. The outcome of this procedure is a ranked list of symbolic expressions that effectively capture the dependence of the exclusion variable on the pMSSM parameters. The top-ranked expression from this process was selected for integration into the global fit pipeline. This result was achieved through six iterative runs of \texttt{Feyn}, exploring a total of approximately $10^7$ symbolic expression models over roughly 840 CPU core-hours.

\begin{figure*}
    \centering
    \includegraphics[scale=0.41]{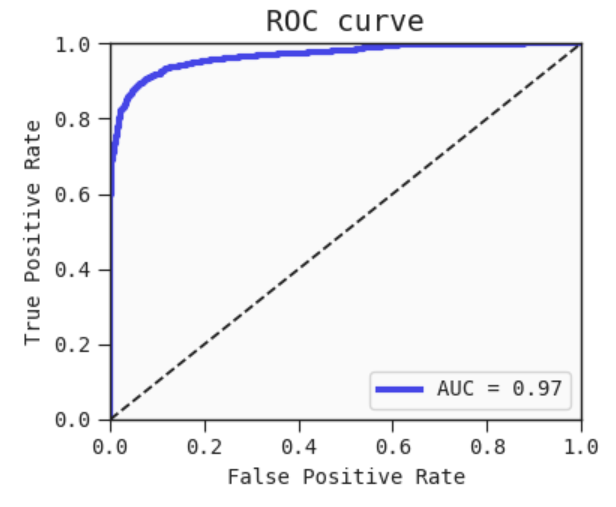}
    \includegraphics[scale=0.41]{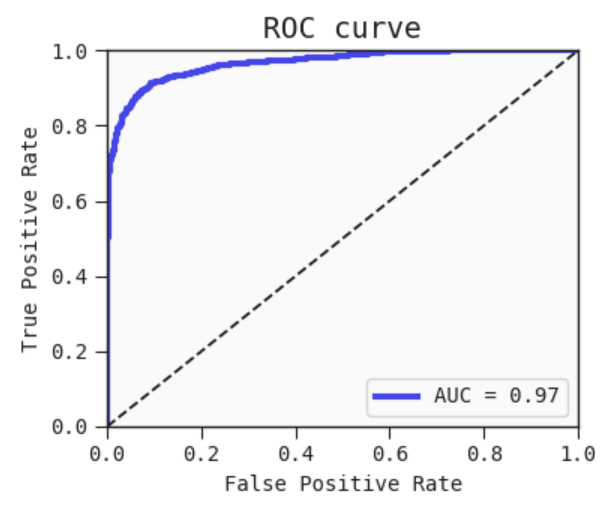}
    \caption{The ROC performance of the symbolic expression learned from the ATLAS dataset showing the ROC curve and AUC values for the training set (left) and test set (right).}
    \label{fig:performance}
\end{figure*}

The performance of the mathematical expression extracted has been evaluated based on receiver operating characteristic (ROC) curve where the True Positive Rate (TPR) against the False Positive Rate (FPR) are plotted. The area under the curve (AUC) achieved by the derived expression is about 0.97 as shown in Figure~\ref{fig:performance}. This signifies excellent discriminative ability, suggesting that the discovered analytical expression is highly effective in identifying positive cases and flagging negative cases.

The symbolic classifier learned by \texttt{Feyn} takes a set of pMSSM parameters $x$ as input and outputs a continuous value between 0 and 1. This value represents a score, or pseudo-probability, indicating the likelihood that a given model belongs to the 'allowed' class by LHC constraints. To convert this continuous score into a binary classification, we employ a Youden threshold of 0.506, empirically derived from the ATLAS dataset.

\section{LHC limits online global fits}
\label{sec:globalFit}
In this section, we demonstrate the application of the symbolic classifier for performing a global fit within the pMSSM framework. The data $d$ utilized for the global fit comprises two categories: $d_{LHC}$, which represents the primary focus of this article, and $d_0$. $d_{LHC}$ consists of the ATLAS limits derived from LHC Run-2 data. The complete dataset is expressed as a direct sum
\be d = d_0 \, \oplus \, d_{LHC}. \ee

The set of measurements constituting $d_0$ for this work includes $\{ \mu_i, \sigma_i \}_{i=1, 2, 3}$, which correspond to the experimental central values and their associated uncertainties, shown in Table~\ref{tab:obs}: the supersymmetric contribution to the muon anomalous magnetic moment, $\delta (g-2)_{\mu}$~\cite{Stockinger:2006zn, Cao:2024axg,2507.09289}, the neutralino cold dark matter relic density, $\Omega_{CDM}h^2$~\cite{Planck:2019nip}, and the Higgs boson mass, $m_H$~\cite{ATLAS:2023dak,CMS:2024eka}. While the primary emphasis of this work is on showcasing how symbolic regression can replace computationally intensive calculations for deriving LHC limits with concise mathematical formulas in global fits, we include these three observables to emulate a comprehensive global fit that combines LHC constraints with other measurements. 

The $d_{LHC}$ component of the dataset consists of limits from ATLAS, represented by the symbolic expression derived in Section~\ref{sec:lhcSymbExpress}. For any given set of model parameters, these mathematical formulas yield a flag indicating compatibility or exclusion based on the learned ATLAS limits. A summary of all measurements and limits considered for the global fit is presented in Table~\ref{tab:obs}.

\begin{table}
\begin{center}{\begin{tabular}{|lll|}
\hline
Observable & Constraint  & References\\ 
\hline
$\delta (g-2)_\mu $ & $(24.9 \pm 4.8) \times 10^{-10}$ & \cite{2507.09289}  \\ 
$\Omega_{CDM} h^2$ & $0.12 \pm 0.02$ & \cite{Planck:2019nip}\\ 
$m_H$ & $125.04 \pm  0.12 \, \gev$ & \cite{ATLAS:2023dak,CMS:2024eka} \\
$d_{LHC}$ & allowed/excluded limits & \cite{1908.08215, 2106.01676, 2108.07586, 2204.13072, 1909.09226, 2103.11684, 1911.12606, 2201.02472} used in \cite{ATLAS:2024qmx} \\  
\hline
\end{tabular}}\end{center}
\caption{Summary for the central values and errors for the second category of data to be used together with LHC limits applied online the global fit of the pMSSM.}
\label{tab:obs}
\end{table}

The parameters of the pMSSM to be fit against the data $d$ are denoted by $\theta$:
\be \label{params}
\theta = \{ M_{1,2,3}; m^{3rd \, gen}_{\tilde{f}_{Q,U,D,L,E}}, m^{1st/2nd \, gen}_{\tilde{f}_{Q,U,D,L,E}}; A_{t,b,\tau}; m_A, \mu, \tan \beta \},
\ee
where $M_1$, $M_2$, and $M_3$ are the gaugino mass parameters, and $m_{\tilde{f}}$ represents the sfermion mass parameters. The parameters $A_{t,b,\tau}$ denote the trilinear scalar couplings. The Higgs sector is defined by the pseudoscalar Higgs mass parameter $m_A$, the mixing parameter $\mu$ that couples the two pMSSM Higgs doublets ($H_1$ and $H_2$), and the ratio of their vacuum expectation values, $\tan \beta = \left<H_2\right>/\left<H_1\right>$.

These parameters are derived from the MSSM supersymmetry breaking parameters assuming, to ensure compatibility with observed CP violation and flavor-changing neutral currents constraints, only real supersymmetry breaking parameters, zero off-diagonal elements in the sfermion mass terms and trilinear couplings, and degenerate first- and second-generation soft supersymmetry-breaking mass terms.

The pMSSM parameter values are sampled from a uniform prior distribution, $\pi(\theta)$, within the ranges specified in Table~\ref{tab:pram_ranges}. Given that the ATLAS analyses under consideration focus on electroweakino production, it is reasonable to fix the mass parameters for first- and second-generation sparticles and third-generation sleptons to a high scale consistent with current LHC limits: $m^{1st/2nd \, gen}_{\tilde{f}_{Q,U,D,L,E}} = m^{3rd \, gen}_{\tilde{f}_{L,E}} = 10 \, \tev$.

The sampling of these parameters is managed by the \texttt{MultiNest}~\cite{Feroz:2007kg} engine, which employs nested sampling algorithm~\cite{Skilling:2004pqw, Skilling:2006gxv, Ashton:2022grj}. The points in the pMSSM parameter space are organized in the Supersymmetry Les Houches Accord (SLHA) format~\cite{Skands:2003cj}. These points are then processed by spectrum generator \texttt{SPHENO}~\cite{Porod:2003um} and \texttt{FeynHiggs}~\cite{Bahl:2018qog} for calculating the Higgs boson mass. The pMSSM points are discarded if they are unphysical due to the absence of electroweak symmetry breaking in the MSSM scalar potential, presence of a tachyonic state within sparticle spectrum, or the LSP not being the lightest neutralino.  

\begin{table}[t]
\centering
\begin{tabular}{ l| r| r }
\hline
Parameter & Range lower bound & Range upper bound  \\
\hline
$M_{1,2}$                          & $-2$ $\tev$  & 2 $\tev$  \\
$M_3$                             & 1 $\tev$  & 5 $\tev$  \\
\hline
$m^{3rd \, gen}_{\tilde{f}_{Q,U,D}}$    & 2 $\tev$  & 5 $\tev$  \\
\hline
$A_t$                            & $-8$ $\tev$  & 8 $\tev$  \\
$A_{b,\tau}$                   & $-2$ $\tev$  & 2 $\tev$  \\
\hline
$m_A$                        & 0 $\tev$  & 5 $\tev$  \\
$\mu$                        & $-2$ $\tev$  & 2 $\tev$  \\
$\tan \beta$            & 1 & 60  \\
\hline
\end{tabular}
\caption{The pMSSM parameter ranges for the global fit. The other parameters in equation \ref{params} not shown in this Table are fixed at $10\, TeV$ following the analyses from which the ATLAS limits were extracted~\cite{ATLAS:2024qmx}: $m^{1st/2nd \, gen}_{\tilde{f}_{Q,U,D,L,E}} = m^{3rd \, gen}_{\tilde{f}_{L,E}} = 10 \, \tev$.}
\label{tab:pram_ranges}
\end{table}

From the computed sparticle spectrum, the prediction for the Higgs boson mass, $m_H$, is obtained. Corresponding predictions for $\delta (g-2)_{\mu}$ and $\Omega_{CDM}h^2$ are calculated using \texttt{micrOMEGAs}~\cite{2312.14894}. For the flags to be compared with the ATLAS limits, predictions are computed using a mathematical formula derived via symbolic classification, as detailed in Section~\ref{sec:lhcSymbExpress}. The set of predictions for each pMSSM parameter space point $\theta$ is given by 
\be
O = \{ m_H,\; \delta (g-2)_{\mu},\; \Omega_{CDM}h^2,\; y(\theta) \}. 
\ee
The likelihood function $L(\theta)$ is then computed as  
\be
\label{like}
L(\theta) = L(d_{LHC}|\theta) \, \prod_{i=1}^{3} \frac{1}{\sqrt{2\pi \sigma_i^2}}\exp\left[- \frac{(O_i - \mu_i)^2}{2\sigma_i^2}\right].
\ee
Here $\mu_i$ and $\sigma_i$ represent the experimental central values and errors for the measurements presented in Table~\ref{tab:obs}. The term $L(d_{LHC}|\theta)$ is defined as 1 if the learned formula $y(\theta)$ flags the parameter point as allowed; otherwise, $L(d_{LHC}|\theta) = 0$. This way, $L(d_{LHC}|\theta)$ enters as a step function derived from the most sensitive ATLAS limits, and therefore the posterior would, in isolation, be interpreted as a prior distribution truncated by a hard exclusion rather than as a smoothly likelihood-driven inference. This conclusion holds only if the three other contributions are neglected; including them restores additional likelihood information and softens the purely cut-based interpretation.

Utilizing the Bayesian inference elements described, we performed two global fits of the pMSSM parameters to the physical observables using \texttt{MultiNest} with 12,000 live points. The first fit was conducted without inclusion of the ATLAS limits, while the second incorporated these limits in the form of symbolic expressions. The resulting posterior samples are employed to infer the impact of the LHC limits on the pMSSM parameters and their correlations, particularly in light of the naturalness line~\cite{AbdusSalam:2015uba} given by 
\be \label{naturalness}
m_A \sim \frac{1}{\sqrt{2}} \, m_Z \, \tan \beta.
\ee

\section{Results and discussion} \label{sec.results}
\begin{figure*}[!tb]
  \centering  
  \includegraphics[width=0.65\textwidth]{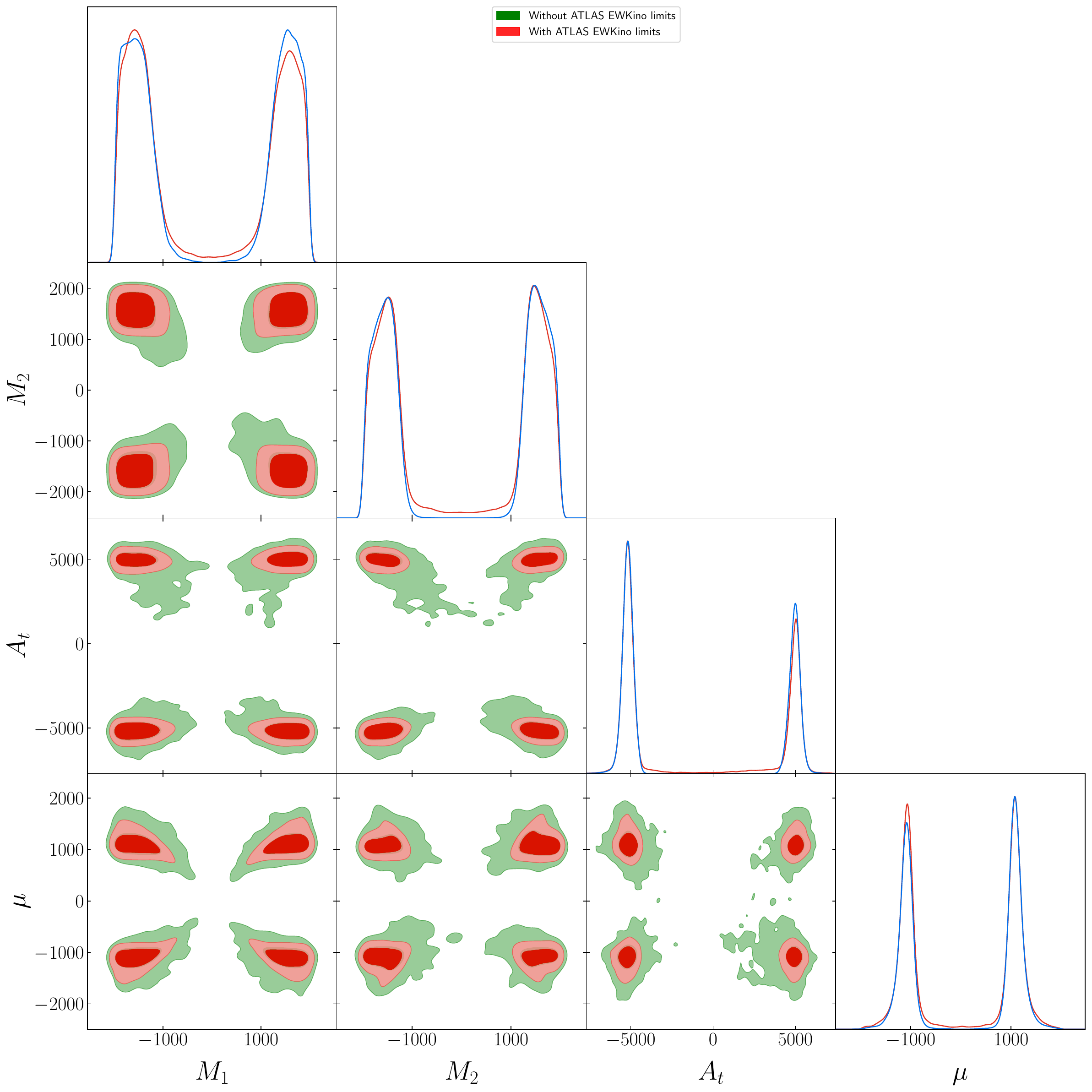}
  \caption{Two-dimensional marginalized posterior distributions of the pMSSM parameters most relevant for the electroweakino scenario targeted by the ATLAS search channels. As indicated in the legend, the distributions in green represent the posterior without the ATLAS limits, while those in red are obtained with the ATLAS limit incorporated during ('online') the global fit. The contours on the two-dimensional plots show the $68\%$ and $95\%$ credibility regions. Red, orange-like, and green filled contours respectively denote the $68\%$ with-ATLAS, $95\%$ with-ATLAS, and $95\%$ without-ATLAS posterior credibility regions. The $68\%$ without-ATLAS posterior credibility region is fully covered by other contours and is therefore not visible in the plots. The ATLAS limits narrowed the posterior from the outer green to the outer red contour lines. This impact can also be seen on the tails of the corresponding one-dimensional distributions. Here, all masses are in GeV natural units. \label{fig:ewkino_params_posterior}}
\end{figure*}

With the two distinct pMSSM global fits -- one incorporating the ATLAS limits approximated by symbolic classification formular, and the other without -- the impact of the LHC limits on the pMSSM parameter space can be analysed. For instance, Figure~\ref{fig:ewkino_params_posterior} presents the two-dimensional marginalized posterior distributions for the pMSSM parameters most relevant to the pMSSM electroweakino scenarios ($M_1, M_2, \mu, A_t$) targeted by the ATLAS search channels. This figures shows the impact of the LHC limits which pushes the posteriors into tighter corners in parameter space. 

\begin{figure*}
  \centering  
  \includegraphics[width=0.6\textwidth]{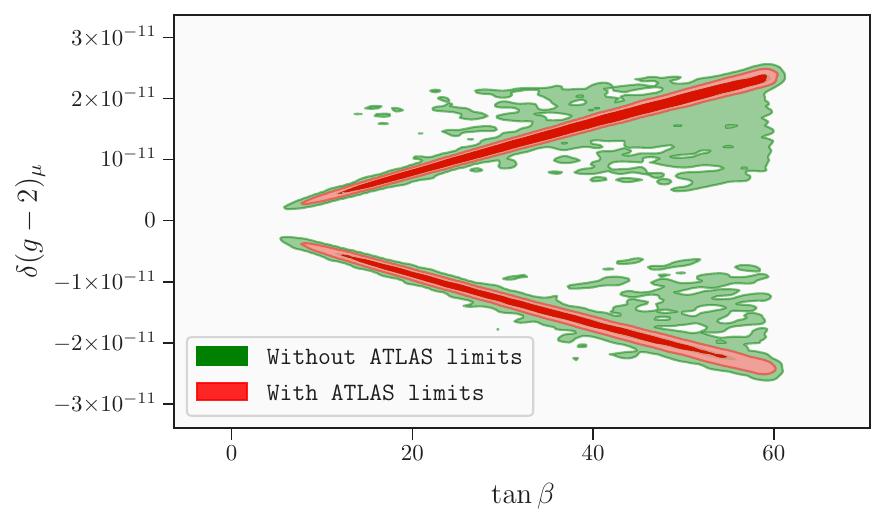}
  \caption{Posterior distributions showing the impact of the ATLAS electroweakino limits on the assumed supersymmetry contribution to the anomalous magnetic moment of the muon $ \delta(g-2)_\mu$ in correlation with the pMSSM parameters, $\tan \beta$. The contours on the two-dimensional plots show the $68\%$ and $95\%$ credibility regions. Red, orange-like, and green filled contours respectively denote the $68\%$ with-ATLAS, $95\%$ with-ATLAS, and $95\%$ without-ATLAS posterior credibility regions. The $68\%$ without-ATLAS posterior credibility region is fully covered by other contours and is therefore not visible in the plots. The ATLAS limits narrowed the posterior from the outer green to the outer red contour lines.\label{fig:tb_gminus2}}
\end{figure*}

Beyond the base parameters, the impact of the LHC limits can be analyse in correlation with other observables. For example, here we consider the the supersymmetry contribution to the anomalous magnetic moment of the muon and also the discuss about electroweak naturalness. This is interesting because light electroweakino states in loops affect $\delta(g-2)_\mu$ and can correlate with the naturalness line condition, $m_A \sim \frac{1}{\sqrt{2}} \, m_Z \, \tan \beta$, via the pMSSM parameters $\tan \beta$ and $m_A$. In Figure~\ref{fig:tb_gminus2}, we show the two-dimensional marginalised posterior distributions in $(\tan \beta, \delta(g-2)_\mu)$ plane. This reveals a clear impact of the ATLAS limits. Specifically, the direct proportionality $\delta(g-2)_\mu \propto \tan \beta$ is well pronounsed for the posterior obtained with the ATLAS limits implemented in the global fit. In contrast, for the posterior without the ATLAS limit, this theoretically derived relation is not valid in the moderate to high $\tan \beta$ regions where the direct proportionality ceazes to be true.

\begin{figure*}
  \centering
    \includegraphics[scale=0.448]{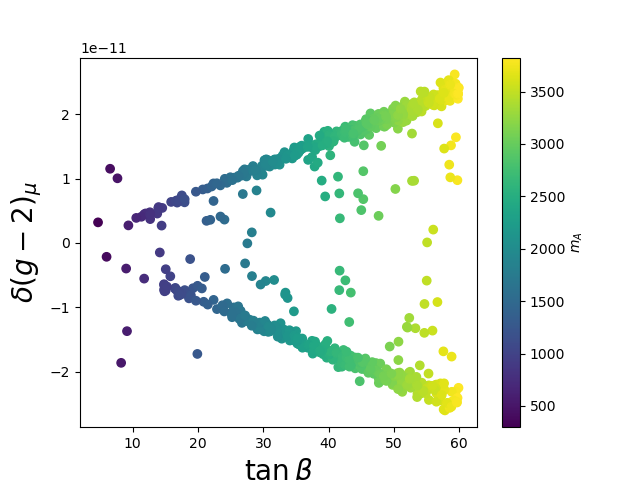}  
    \includegraphics[scale=0.448]{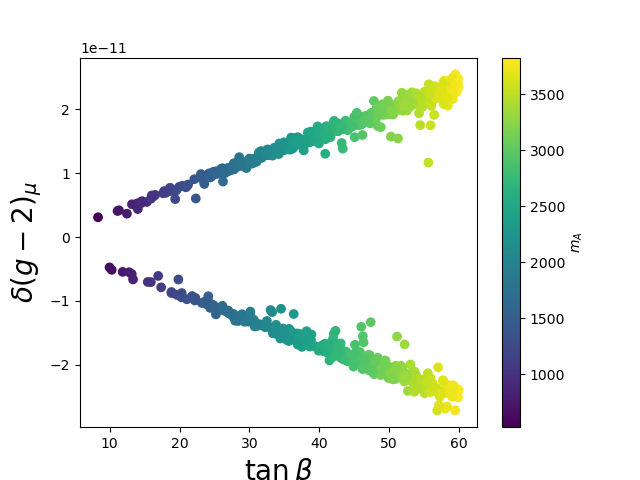}
    \caption{3D Scatter plots of the pMSSM posterior showing the ATLAS limit versus naturalness line cut implications. The left-side (right-side) plot is for the posterior sample without (with) the ATLAS limits included during the global fits. For both plots, the naturalness line cut, $m_A/ \tan \beta \sim \frac{1}{\sqrt{2}}$, has been applied. }
    \label{fig:natural}
\end{figure*}

The correlation with $\tan \beta$ strongly links the supersymmetry contribution $\delta(g-2)_\mu$ to the pMSSM parameter $m_A$. Figure~\ref{fig:natural} shows that the three-dimensional scatter plots of the posterior samples with the naturalness line cute ($m_A/ \tan \beta \sim \frac{1}{\sqrt{2}}$) applied. Again, in this plot also, it can be seen that the anomalous posterior points that tends to spoil the relationship $\delta(g-2)_\mu \propto \tan \beta$ relatively get cleared out for the case of the global fit with the ATLAS limits applied.

From the posterior samples, there are indications that the ATLAS limits and 'naturalness line' are squeezing the pMSSM parameters in opposite directions into tighter corners when viewed across the supersymmetry contribution to the anomalous magnetic moment of the muon perspective. With the naturalness line cut, $m_A$ must be less than $4 \, \tev$ for $\tan \beta \in [1, 60]$. With ATLAS limit and naturalness line applied, $m_A$ must be greater than $500 \, \gev$ or $100 \, \gev$ without the ATLAS limit.

For a given range of $\tan \beta$ values, the maximum allowed $m_A$ can be computed from the naturalness line. Concurrently, the ATLAS limit discussed in this letter drives the lower limit on $m_A$ to higher values. The interplay between these LHC constraints and the naturalness line offers valuable insights into the potential for the LHC to exclude supersymmetry in the absence of a discovery. This is because these two constraints can constrain the parameter space in opposing directions, effectively reducing it to smaller regions. Thus, it is indeed possible that "the LHC can rule out the MSSM"~\cite{AbdusSalam:2011hd}.

\section{Conclusions and outlook} \label{sec.conclusion}
In this work, we demonstrated that experimental exclusion boundaries can be encoded as explicit mathematical expressions via symbolic classification. This strategy enables the inclusion of LHC constraints in pMSSM global fits during the scanning stage, in contrast to the traditional approach of \emph{post-processing} posterior samples. The latter typically requires repeated, computationally expensive evaluation of collider limits (including event-level recast or likelihood computations), and can therefore become prohibitively slow for high-dimensional global inference. By replacing these costs with fast, analytic approximations, the proposed framework provides an efficient route to incorporate LHC limits ``online'' global fits analyses.

At the methodological level, symbolic regression provides a principled way to translate a set of discrete ATLAS exclusion points considered into compact formulas that straightforward to embed in probabilistic pipelines. We trained a symbolic classifier using ATLAS inputs for the pMSSM electroweakino scenario. We find that the resulting classifier is reliable, achieving an area under curve of the receiver operating characteristic curve of approximately $97\%$. We then used the learned symbolic expression within a dedicated pMSSM framework to illustrate how LHC limits can be incorporated \emph{online} global fits. Here ``online'' refers to the direct application of limits during the generation of posterior samples rather than post-processing. This approach bypasses the need for the traditional, time-consuming collider-limit evaluation for each sampled parameter point, thereby making the computational burden compatible with global fit scans. To our knowledge, this constitutes the first demonstration of LHC limits being included online an pMSSM global fit in this manner. 

Beyond the computational advantage, we also highlighted the physical value of such an approach. Global fits consolidate complementary experimental information into a coherent probabilistic picture of the BSM parameter space, clarifying which regions are supported or disfavoured and guiding where experimental and theoretical efforts should be focused. Applying our method to the ATLAS-constrained electroweakino pMSSM, we showed that the impact of LHC constraints can be traced not only in the marginal posteriors of model parameters but also in their correlations with additional observables. For example, the interplay between the supersymmetry contribution to the muon anomalous magnetic moment $\delta(g-2)_\mu$ and electroweak naturalness, characterized by the naturalness-line condition, $m_A/ \tan \beta \sim \frac{1}{\sqrt{2}}$, becomes sharper once ATLAS constraints are imposed.

This study is a proof of principle: the framework is general and can, in principle, be extended to other BSM models and other LHC analyses, offering a practical pathway to integrate collider information efficiently into global fits for particle physics phenomenology. In the original ATLAS dataset, there could be few problematic points with bugs related to the decay of neutralino NLSP to LSP plus photon and the decay of lightest chargino to LSP plus pion. The removal of such points will not change the conclusions and message of this paper.

Looking ahead, the posterior distributions for the lightest stop and sbottom masses indicate that the current LHC direct reach has not yet probed the stop-mass region suggested by the Higgs-boson mass. Future LHC runs and the HL-LHC can substantially improve this situation by accessing this portion of the parameter space.

Finally, the possibility of extracting combined LHC likelihoods in~\texttt{SModelS}~\cite{MahdiAltakach:2023bdn,Altakach:2023tsd,Constantin:2025bqp} provides access to global LHC-limits based likelihoods beyond simple step functions as LHC constraints. An interesting line of ongoing work following the research presented here is the extraction of global LHC likelihoods available through \texttt{SModelS}~\cite{2312.16635,2505.11251}. This will substantially improve global fits by replacing the current binary allowed-versus-excluded classification with a full-fledged LHC likelihood framework.

\paragraph*{Acknowledgement:}
I would like to thank my collaborators on symbolic regression methods, Steven Abel, Miguel Crispim Romão, and Deaglan Bartlett for discussions. Many thanks to Roberto Franceschini and Sabine Kraml for valuable comments on the draft of the paper. This work is based upon research funded by Iran National Science Foundation (INSF) under project No. 4040025.

\section*{Data Availability}
The data that support the findings of this article are openly available~\cite{abdussalam_2026_21262199}.

\bibliographystyle{inspire}
\bibliography{references}  

\end{document}